\documentclass[aps,preprint,prd,showpacs,nofootinbib]{revtex4}
\usepackage{amsmath}
\usepackage{graphicx}
\usepackage{dcolumn}
\usepackage{bm}
\usepackage{amssymb}
\usepackage{latexsym}
\usepackage{color}

\def\be{\begin{equation}}
\def\ee{\end{equation}}
\def\ba{\begin{eqnarray}}
\def\ea{\end{eqnarray}}

\begin{document}

\title{Pre-inflationary genesis with CMB B-mode polarization    }

\author{Zhi-Guo Liu$^{1}$\footnote{Email: liuzhiguo08@mails.ucas.ac.cn}}
\author{Hong Li$^{2,3}$\footnote{Email: hongli@ihep.ac.cn}}
\author{Yun-Song Piao$^{1,4}$\footnote{Email: yspiao@ucas.ac.cn}}

\affiliation{$^1$ School of Physics, University of Chinese Academy
of Sciences, Beijing 100049, P.R.China}

\affiliation{$^2$ Key Laboratory of Particle Astrophysics,
Institute of High Energy Physics, Chinese Academy of Science,
P.O.Box 918-4, Beijing 100049, P.R.China}

\affiliation{$^3$ National Astronomical Observatories, Chinese
Academy of Sciences, Beijing 100012, P.R.China}

\affiliation{$^4$ State Key Laboratory of Theoretical Physics, Institute of Theoretical Physics, \\
Chinese Academy of Sciences, P.O. Box 2735, Beijing 100190, China}

\begin{abstract}


Recent B-mode polarization observation seems to imply the tensor
tilt $n_T\gtrsim 1$ at large angular scale, if the primordial
signal is dominated. We show that for a primordial universe, which
is in a slowly expanding genesis phase before the slow-roll
inflation, the primordial tensor spectrum will get a large-scale
cutoff, i.e. $n_T\gtrsim 1$ at large scales while $n_T\simeq 0$ at
small scale. We find that this inflationary scenario not only may
be consistent with the observation, but also predicts a
large-scale anomaly in BB power spectrum, i.e. due to the large
suppression of tensor perturbation amplitude we will hardly see
the reionization bump at low-$l$, which may be falsified by the
Planck polarization data.

\end{abstract}

\maketitle

The discovery of the cosmic microwave background (CMB) B-mode
polarization, contributed by the primordial gravitational wave, is
crucial for understanding the physics of the early universe,
especially solidifying our confidence that inflation has ever
occurred. Recently, the BICEP2 collaboration has declared the
detection of the CMB B-mode polarization \cite{Ade:2014xna}.
However, it might be wholly or partly due to polarized dust
emission \cite{Adam:2014bub}, see also \cite{Mortonson:2014bja}.

Though the primordial B-mode polarization could be partially
caused by other sources
\cite{Lizarraga:2014eaa},\cite{Moss:2014cra}, it is
highly-probably induced by the primordial tensor perturbation. The
Planck temperature data have put the constraint $r < 0.11$ (95\%
C.L.) \cite{Ade:2013kta} on the tensor amplitude, but no limit for
the tensor tilt $n_T$. Recent B-mode polarization data, after the
dust is neglected, seem to favor $n_T\gtrsim 1$
e.g.\cite{Li:2014cka},\cite{Smith:2014kka},\cite{Lewis2014},\cite{Gerbino:2014eqa},\cite{Hu:2014aua},\cite{Zhang:2014},
see also \cite{Martin:2014lra} for the analysis for the
consistence of slow-roll inflationary models. However, despite
this result, it is interesting to think over the physics of
inflationary universe relevant with $n_T\gtrsim 1$.

The conventional inflation model requires $n_T=-2\epsilon < 0$, in
which $\epsilon = -{\dot H}/H^2$. $n_T>0$ implies either new
mechanism or non-Bunch-Davis vacuum is involved during inflation,
e.g.\cite{Mukohyama:2014gba},\cite{Cai:2007xr},\cite{Ashoorioon:2013eia},
or the NEC is broken, see e.g.\cite{Wang:2014kqa} for comments. In
past, inflationary model with the NEC violation has been also
proposed \cite{Piao:2004tq},\cite{Baldi:2005gk}. However, since
the almost scale-invariance of the scalar perturbation requires
$|\epsilon|\ll 1$, we generally have $|n_T|\ll 1$ during the
inflation. Thus $n_T\gtrsim 1$ seems be a challenge for the
inflationary scenario. There is an alternative to inflation, in
which the primordial universe is slowly expanding, i.e. genesis
scenario \cite{CNT},\cite{Liu:2011ns},\cite{Hinterbichler:2012fr},
also earlier \cite{Piao:2003ty}, see \cite{Rubakov:2014jja} for a
review. In this scenario, $n_T= 2$, thus its tensor amplitude is
exponentially low at large scale, which is negligible. The case is
similar to that in ekpyrotic scenario \cite{Khoury:2001wf}.

However, if $n_T\gtrsim 1$ is only at large angular scale, while
at intermediate and small scales $n_T\simeq 0$ is satisfied, it
will be possible to put $r$ at large and intermediate scales in a
detectable level. In this case, $n_T\gtrsim 1$ at large angular
scale might be a potential reflection of the pre-inflationary
evolution.

The pre-inflationary evolution may bring signals in TT power
spectrum at large angular scale,
e.g.\cite{Piao:2003zm},\cite{Liu:2013kea},\cite{Dudas:2012vv},
which mainly imprinted by the primordial scalar perturbation.
Besides the evolution of $a$, the scalar spectrum is also affected
by the model parameters associated with the field dynamics.
However, the tensor perturbation spectrum only encodes the
behavior of $a$, which straightly records the evolution of the
primordial universe. Thus $n_T\gtrsim 1$ at large scale will be a
significant hint of what occurring before inflation. The idea of a
short period of inflation has been studied in Ref.\cite{Ramirez},
without a pre-inflationary NEC violation. However, a
pre-inflationary evolution with NEC violation is interesting,
since it helps to solve the initial singularity problem of
inflationary universe \cite{Borde:2001nh}.

Here, we will propose an inflationary scenario, in which the
pre-inflationary evolution of the primordial universe is a slowly
expanding genesis phase. We find that this pre-inflationary
genesis will make the primordial tensor spectrum get a large-scale
cutoff, i.e. $n_T> 1$ at large scale, while $n_T\simeq 0$ at small
scale. Thus this scenario not only may be consistent with the
observation, but also predicts a large-scale anomaly in BB power
spectrum, i.e. due to the large suppression of tensor perturbation
amplitude we will hardly see the reionization bump at $l<10$,
which may be falsified by the upcoming Planck polarization data.

\begin{figure}[t]
\includegraphics[width=8.0cm]{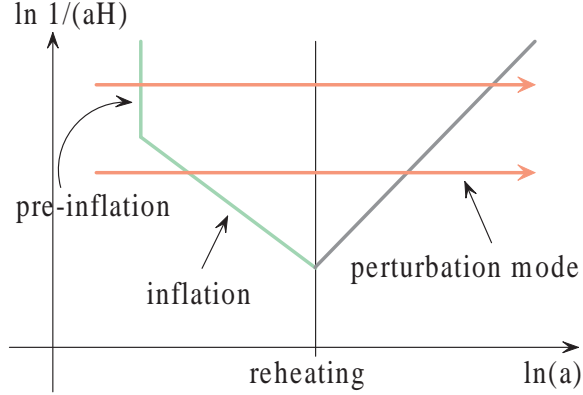}
\caption{The sketch of the evolution of primordial perturbations
in our scenario. The perturbation mode at large scale is from the
pre-inflationary genesis phase, while others are from the
inflationary phase. } \label{fig:perturbation}
\end{figure}

We begin with briefly illustrating the background evolution and
the scalar perturbation. Before the slow-roll inflation, the
universe is in a slowly expanding genesis phase, \be a\sim e^{
 1\over M^{b}(t_*-t)^{b}}, \label{gen}\ee where $b>0$ and $t$ runs from
negative infinite. Initially $|t|\gg
 |t_*|$ and $M^{b}(t_*-t)^{b}\gg 1$,
the universe is slowly expanding
\cite{CNT},\cite{Liu:2011ns},\cite{Hinterbichler:2012fr}, see
\cite{Khoury:2010gw},\cite{Geshnizjani:2011rm} for alternatives.
Here, \be H\sim { 1\over M^{b}(t_*-t)^{b+1}}, \label{H}\ee which
initially may be negligible, but rapidly increases with the time.
Thus $\epsilon=-{\dot H}/H^2\ll -1$ breaks NEC, which seems
implying the ghost instability. However, the ghost instability may
be dispelled by applying Galileon field \cite{NRT}. We may find
initially $|\epsilon|\sim M^{b}(t_*-t)^{b}\gg 1$ and rapidly
decreases with time. When $|\epsilon|<1$, the slow expansion of
universe ends, after this \be H=H_*\sim {1\over |t_*|}\simeq
H_{inf} \label{Hstar}\ee is unchanged and the slow-roll inflation
will begin, in which \be a\sim e^{H_{inf} (t-t_*)}\label{inf}\ee
and $\epsilon_{inf}\ll 1$. We plot Fig.\ref{fig:perturbation} for
illustrating this scenario and the evolution of primordial
perturbations.

In slow-roll inflationary phase, the scalar perturbation is
scale-invariant, i.e. $n_{\cal R}-1\ll 1$. In pre-inflationary
genesis phase, the scalar perturbation has been investigated. In
Refs.\cite{Liu:2011ns},\cite{Piao:2010bi}, $a\sim e^{
 1\over (t_*-t)^{4}}$, which gives the scale-invariant adiabatic
perturbation, \be {\cal P}_{\cal R}={H_*^2\over 8\pi^2 \epsilon_*
M^2_P} \ee with $H_*$ defined in (\ref{Hstar}). In Ref.\cite{CNT},
$a\sim e^{
 1\over (t_*-t)^{2}}$, the adiabatic
perturbation is not scale-invariant, but the perturbations of
conformal light scalar fields may be scale invariant
\cite{Rubakov:2009np},\cite{CNT},\cite{Hinterbichler:2011qk},
which will contribute the curbature perturbation.
Thus the pre-inflationary scalar spectrum may be naturally
coincident with that of slow-roll inflation, in spite of the
tensor perturbation. Here, since BB is contributed only by the
tensor perturbation, we will not involve the scalar perturbation,
and only require that it is consistent with all data. We will
clarify the details of the model building elsewhere.

However, the pre-inflationary tensor spectrum can hardly be the
same with that of slow-roll inflation. We will investigate the
primordial tensor perturbation in this scenario. The action of the
tensor perturbation $h_{ij}$ is \be S_2\sim \int d\eta d^3x
{a^2M_P^2\over 4}\left({{h_{ij}}^\prime}^2-(\partial
{h_{ij}})^2\right). \label{h}\ee The equation of $h_{ij}$ in
momentum space is  \be v_k^{\prime\prime}
+\left(k^2-{a^{\prime\prime}\over a}\right) v_k = 0, \label{uk}\ee
after $v_k \equiv aM_P{h_{ij,}}_k/\sqrt{2}$ is defined, where $'$
is the derivative with respect to conformal time $\eta=\int dt/a$.


We, for simplicity, will adopt an instantaneous matching of
backgrounds (\ref{gen}) and (\ref{inf}) at $\eta=0$, and have
\ba a & \simeq & a_*,
\,\,\,for\,\,{\rm preinflationary  }\,\,{\rm genesis}\,\,{\rm phase}, \label{leq} \nonumber\\
& & {a_*\over 1-{\cal H}_*\eta}~, \,\,\, for \,\,{\rm
inflationary}\,\,{\rm phase}, \label{geq} \ea respectively, ${\cal
H}_*$ is the comoving $H_*$ at $\eta=0$, which sets the slow-roll
inflationary scale by $H_{inf}=H_*={\cal H}_*/a_*$.


When $k^2\gg {a^{\prime\prime}\over a}$, i.e. the perturbation is
deeply inside its horizon, $v_k$ oscillates with a constant
amplitude, \be v_k\sim {1\over \sqrt{2k}} e^{-ik\eta}.
\label{ini}\ee When $k^2\ll {a^{\prime\prime}\over a}$, i.e. the
perturbation is far outside the horizon, in the genesis phase the
solution of Eq.(\ref{uk}) is \ba
v_k=\sqrt{\frac{\pi}{4}x}H^{(1)}_{1/2}(x), \ea where $x={k/{\cal
H}_*}-k\eta$, $H^{(1)}_{1/2}$ is the 1/2th order Hankel function
of the first kind,
while in the inflationary phase the solution of Eq.(\ref{uk}) is
\ba & &{v_k} =\sqrt{x}\, c_1 H_{3/2}^{(1)}(x)+\sqrt{x}\,c_2
H_{3/2}^{(2)}(x), \label{vki} \ea where $H_{3/2}^{(1)}$ and
$H_{3/2}^{(2)}$ are the 3/2th-order Hankel function of the first
kind and the second kind, respectively, $c_1$ and $c_2$ are only
dependent on $k$.

\begin{figure}[htbp]
\includegraphics[scale=0.6,width=6.0cm]{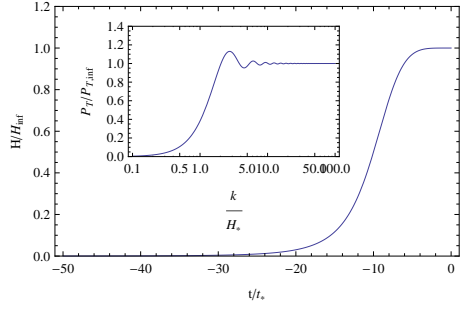}
\caption{The sketch of the evolution of $H$. During the
pre-inflationary phase, it initially is negligible, but rapidly
increases with the time. After $H=H_{inf}$, it becomes nearly
unchanged, the slow-roll inflation will begin. The inset is the
power spectrum (\ref{ps}) of the primordial tensor perturbation in
our scenario with respect to $k/{\cal H}_*$, which has a
large-scale cutoff, i.e. $n_T\gtrsim 1$ for $k< {\cal H}_*$ and
$n_T\simeq 0$ for $k> {\cal H}_*$. The figure of the evolution of
$H$ with the time should be almost coincident with that of the
tensor spectrum with respect to $k$, because the amplitude of
tensor mode with wavelength $1/k$ is $\sim H$ at the time $k=aH$.}
\end{figure}

\begin{figure}[t]
\begin{center}
\includegraphics[scale=0.6,width=8.0cm]{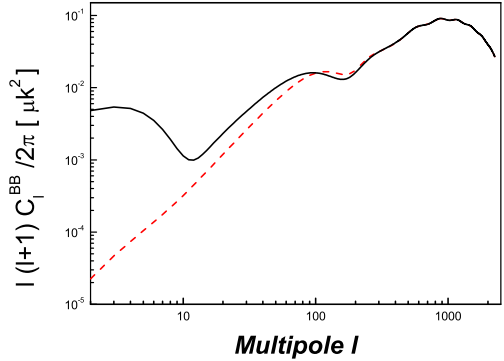}
\caption{Theoretical CMB BB power spectra for the conventional
slow-roll inflationary scenario (black solid line), in which the
tensor spectrum is power-law, and $\Omega_bh^2=0.02184$,
$\Omega_ch^2=0.118$, $\tau=0.08$, $100\Theta_s=1.04$, $n_{\cal
R}=0.96$ $A_s=2.14\times10^{-9}$, and our scenario (red dashed
line), in which $n_T$ is replaced with Eq.(\ref{ps}).
\label{figure1:bestCMB}}
\end{center}
\end{figure}

The continuity of $h_{ij}$ gives $c_1$ and $c_2$. Thus the
spectrum of $h_{ij}$ is ${\cal P}_{T} = {2k^3\over
\pi^2}\left|{v_k\over a}\right|^2$, which equals to \ba {\cal
P}_{T}& = & {4{H}_{inf}^2\over \pi^3 M_P^2}k\left|c_1
-c_2\right|^2
\nonumber\\ & = & {\cal P}_{T,inf} \left(1+\frac{\cos{2{\tilde
k}}}{{\tilde k}^2}-\frac{\sin{2{\tilde k}}}{{\tilde
k}^3}+\frac{\sin^2{\tilde k}}{{\tilde k}^4}\right), \label{ps} \ea
where ${\cal P}_{T,inf}={2{H}_{inf}^2\over \pi^2 M_P^2}$ is that
of the slow-roll inflation, and ${\tilde k}=k/{\cal H}_*$ is
defined for simplicity. When ${\tilde k}\ll 1$, i.e. $k\ll {\cal
H}_*$, we can have approximately \ba {\cal P}_{T} & \sim & {\tilde
k}^2,\label{P1}\ea where all terms with $1/{\tilde k}^2$, as well
as ${\tilde k}^0$, just cancel, which is that in the genesis phase
with $n_T=2$ \cite{CNT}. While for ${\tilde k} \gg 1$, \be {\cal
P}_{T}\simeq {\cal P}_{T,inf}, \ee which is that of slow-roll
inflation. The result is consistent with the evolutions of the
perturbation modes plotted in Fig.\ref{fig:perturbation}. However,
for the mode with ${\tilde k}\sim 1$, i.e. $1/k\sim 1/{\cal H}_*$
is around the comoving cutoff scale, we have $n_T\sim 1$. We plot
${\cal P}_{T}(k)$ in the inset of Fig.2, in which for $k> {\cal
H}_*$, the spectrum is almost scale-invariant with a decaying
oscillation, and for $k<{\cal H}_*$ it gets a cutoff.


We modify the Boltzmann code CAMB\cite{camb} to calculate the
lensed BB power spectrum, which is plotted in
Fig.\ref{figure1:bestCMB}.
We set the comoving cutoff scale $1/{\cal H}_*$ at $l\sim 60$, at
which $n_T\sim 1$, and see that the large-scale cutoff of the
primordial tensor spectrum (\ref{ps}) brings a BB power
suppression.

Noting that $n_T\gtrsim 1$ is conflicted with the observations of
the pulsar timing, BBN, LIGO at smaller scales, which put
$n_T<0.4$ \cite{Smith:2014kka}, see also \cite{Gerbino:2014eqa}.
Here, interestingly, we have $n_T>1$ at large scale, but
$n_T\simeq 0$ at small scale, which makes our scenario also comply
with the constraints at small scale well.

There is generally a recombination peak at $l\sim 80$ and a
reionization bump at $l<10$ in $\Lambda$CDM $+~r$ power-law model.
Both are the imprints of the primordial tensor perturbation. The
detection of the recombination peak will be a confirmation that
inflation has ever occurred, while Planck with all-sky coverage
will highly-probably detect the reionization bump at low-$l$,
which might help us to understand the physics of pre-inflationary
universe. The amplitude and shape of the reionization bump record
the reionization history of universe. However, since we have
$n_T=2$ at larger scale, the perturbation modes at low-$l$ will
acquire a larger suppression, which makes the reionization bump in
our BB power spectrum hardly visible. The upcoming Planck
polarization data may falsify our scenario. The effect of
$n_T\gtrsim 1$ on the reionization bump in BB power spectrum is
estimated in Appendix. In addition, the large reduction of tensor
amplitude at large scale also helps to reduce power in TT power
spectrum at low-$l$, where Planck showed a deficit.

Here, we set the comoving cutoff scale $1/{\cal H}_*$ at $l\sim
60$. It should be mentioned that if $1/{\cal H}_*$ is at $l\sim
10$, the suppression of the reionization bump in BB power spectrum
will be weaken, a low bump will appear, like in
Ref.\cite{Wang:2014abh}. We will study such a case elsewhere.

In summary, we have proposed an inflationary scenario, in which
the pre-inflationary evolution of the primordial universe is a
slowly expanding genesis phase. Predictably, with this scenario,
we will not see the reionization bump in BB power spectrum at
$l<10$, which will highly-probably detected by Planck. The
pre-inflationary genesis requires the dramatic violation of NEC,
which might have potential implications,
e.g.\cite{Rubakov:2013kaa},\cite{Liu:2013xt},\cite{Elder:2013gya},
the upcoming Planck polarization data would tell whether it has
ever occurred.


\textbf{Acknowledgments}

This work was talked in "Advanced Seminar" on Apr.16, organized by
Jianxin Lu. We thank Yungui Gong, Qing-Guo Huang, Mingzhe Li for
discussions during the seminar. YSP is supported by NSFC, No.
11222546, and National Basic Research Program of China,
No.2010CB832804. HL is supported by NSFC, No. 11322325 and
11033005.

\section*{Appendix: The parameterization of tensor spectrum with large-scale cutoff }


We will estimate the effect of $n_T\gtrsim 1$ on the reionization
bump in BB power spectrum at $l<10$. ${\cal P}_T$ may be
parameterized as \be {\cal P}_T = {{\cal P}_{T,inf}\over
1+A_*\left({{\cal H}_*}\over k\right)^{n_{cutoff}}},
\label{ps1}\ee where ${\cal P}_{T,inf}$ is that of the slow-roll
inflation, both $A_*$ and $n_{cutoff}$ are parameters reflecting
the shape of ${\cal P}_T$. We have $n_T \simeq 0$ for $k\gg {\cal
H}_*$ and $n_T\simeq n_{cutoff} $ for $k\ll {\cal H}_*$. Thus the
value of $n_{cutoff}$ sets a large-scale cutoff.

When $n_{cutoff}=2$, the spectrum (\ref{ps1}) may simulate
(\ref{ps}) with a negligible oscillation in the regime $k>{\cal
H}_*$. When $n_{cutoff}=1$, it is that in Ref.\cite{Liu:2013iha},
in which the pre-inflationary expanding phase is a
superinflationary evolution \be a\sim {1\over H_*(t_*-t)} \ee and
$\epsilon \simeq -1$. Here, the superinflation is defined as the
evolution with ${\dot H}/H^2>0$. Thus the slowly expanding genesis
also belongs to the superinflation, but since $\epsilon\ll -1$,
the expansion is actually slow.

It is validated in Ref.\cite{Piao:2004jg} that if the
pre-inflationary universe is expanding, $n_T\leqslant 2$, and if
it is contracting, $n_T\leqslant 3$. Thus for $n_{cutoff}=3$, the
pre-inflationary universe should be contracting, it may be that in
bounce inflation scenario \cite{Piao:2003zm},\cite{Liu:2013kea},
in which before the slow-roll inflation the universe is in a
contracting phase with $a\sim (t_*-t)^{1/3}$, on which a cyclic
universe may be based \cite{Piao:2004me}. Here, the same cutoff
also appears in scalar spectrum. However, since the comoving
cutoff scale for the scalar spectrum is hardly extended to $l>10$,
or the model is not consistent with Planck temperature data, thus
the reionization bump of BB power spectrum in bounce inflation
will still exist but slightly lower \cite{Wang:2014abh}, see also
\cite{Xia:2014tda},\cite{Qiu:2014nla} for other pre-inflationary
contractions.

\begin{figure}[t]
\begin{center}
\includegraphics[scale=0.6,width=8.0cm]{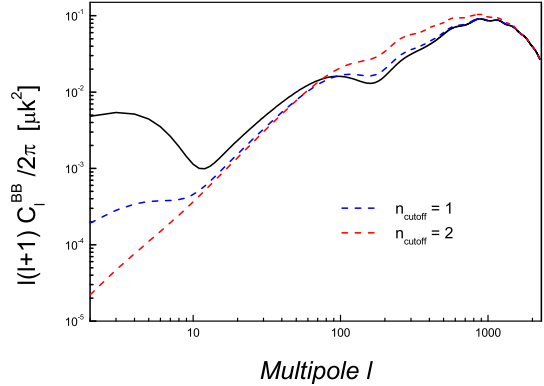}
\caption{Theoretical CMB BB power spectra for the conventional
slow-roll inflationary scenario (black solid line), in which the
tensor spectrum is power-law, and $\Omega_bh^2=0.02184$,
$\Omega_ch^2=0.118$, $\tau=0.08$, $100\Theta_s=1.04$, $n_{\cal
R}=0.96$ $A_s=2.14\times10^{-9}$, and our parameterized model, in
which $n_T$ is replaced with the parameterization (\ref{ps1}). We
consider two cases: $n_{cutoff} = 1$ (blue dashed line) and
$n_{cutoff} = 2$ (red dashed line).\label{figure2:bestCMB}}
\end{center}
\end{figure}

We plot the effect of $n_{cutoff}\gtrsim 1$ on the reionization
bump at $l<10$ in Fig.\ref{figure2:bestCMB}. When $n_{cutoff}=2$,
there is not the reionization bump, since the tilt of ${\cal P}_T$
is large, which leads to a highly large reduction of the
perturbation modes at low-$l$ and makes the reionization bump
hardly visible. When $n_{cutoff}= 1$, the reionization bump will
appear but be compressed compared with that of $n_{cutoff}= 0$.
The parameterization (\ref{ps1}) actually helps to qualitatively
distinguish the inflationary models with different
pre-inflationary evolutions giving different $n_{cutoff}$.

\begin{thebibliography}{99}




\bibitem{Ade:2014xna}
  P.~A.~R.~Ade {\it et al.}  [BICEP2 Collaboration],
  arXiv:1403.3985 [astro-ph.CO].

\bibitem{Adam:2014bub}
  R.~Adam {\it et al.}  [Planck Collaboration],
  arXiv:1409.5738 [astro-ph.CO].

\bibitem{Mortonson:2014bja}
  M.~J.~Mortonson and U.~Seljak,
  arXiv:1405.5857 [astro-ph.CO].


\bibitem{Lizarraga:2014eaa}
  J.~Lizarraga, J.~Urrestilla, D.~Daverio, M.~Hindmarsh, M.~Kunz and A.~R.~Liddle,
  arXiv:1403.4924 [astro-ph.CO].

\bibitem{Moss:2014cra}
  A.~Moss and L.~Pogosian,
  arXiv:1403.6105 [astro-ph.CO].

\bibitem{Ade:2013kta}
  P.~A.~R.~Ade {\it et al.}  [Planck Collaboration],
  arXiv:1303.5075 [astro-ph.CO].



\bibitem{Li:2014cka}
  H.~Li, J.~-Q.~Xia and X.~Zhang,
  arXiv:1404.0238 [astro-ph.CO].

\bibitem{Smith:2014kka}
  K.~M.~Smith, C.~Dvorkin, L.~Boyle, N.~Turok, M.~Halpern, G.~Hinshaw and B.~Gold,
  Phys.\ Rev.\ Lett.\  {\bf 113}, 031301 (2014)  [arXiv:1404.0373 [astro-ph.CO]].

\bibitem{Martin:2014lra}
  J.~Martin, C.~Ringeval, R.~Trotta and V.~Vennin,
  arXiv:1405.7272 [astro-ph.CO].

\bibitem{Lewis2014} A. Lewis,
[http://cosmocoffee.info/viewtopic.php?t=2302].


\bibitem{Gerbino:2014eqa}
  M.~Gerbino, A.~Marchini, L.~Pagano, L.~Salvati, E.~Di Valentino and A.~Melchiorri,
  Phys.\ Rev.\ D {\bf 90}, 047301 (2014)  [arXiv:1403.5732 [astro-ph.CO]].

\bibitem{Hu:2014aua}
  B.~Hu, J.~W.~Hu, Z.~K.~Guo and R.~G.~Cai,
  Phys.\ Rev.\ D {\bf 90}, 023544 (2014)  [arXiv:1404.3690 [astro-ph.CO]].

\bibitem{Zhang:2014}
  Y.~-H. Li, J.~-F. Zhang, X. Zhang, arXiv:1405.0570
  [astro-ph.CO].


\bibitem{Mukohyama:2014gba}
  S.~Mukohyama, R.~Namba, M.~Peloso and G.~Shiu,
  arXiv:1405.0346 [astro-ph.CO].

\bibitem{Cai:2007xr}
  Y.~f.~Cai and Y.~S.~Piao,
  Phys.\ Lett.\ B {\bf 657}, 1 (2007)  [gr-qc/0701114]; 
  Y.~F.~Cai and Y.~Wang,
  Phys.\ Lett.\ B {\bf 735}, 108 (2014)  [arXiv:1404.6672 [astro-ph.CO]].

\bibitem{Ashoorioon:2013eia}
  A.~Ashoorioon, K.~Dimopoulos, M.~M.~Sheikh-Jabbari and G.~Shiu,
  JCAP {\bf 1402}, 025 (2014)  [arXiv:1306.4914 [hep-th]]; 
  Phys.\ Lett.\ B {\bf 737}, 98 (2014)  [arXiv:1403.6099 [hep-th]].

\bibitem{Wang:2014kqa}
  Y.~Wang and W.~Xue,
  arXiv:1403.5817 [astro-ph.CO].



\bibitem{Piao:2004tq}
  Y.~-S.~Piao and Y.~-Z.~Zhang,
  Phys.\ Rev.\ D {\bf 70}, 063513 (2004)  [astro-ph/0401231].

\bibitem{Baldi:2005gk}
  M.~Baldi, F.~Finelli and S.~Matarrese,
  Phys.\ Rev.\ D {\bf 72}, 083504 (2005)  [astro-ph/0505552].




\bibitem{CNT}
  P.~Creminelli, A.~Nicolis and E.~Trincherini,
  JCAP {\bf 1011}, 021 (2010)  [arXiv:1007.0027 [hep-th]].

\bibitem{Liu:2011ns}
  Z.~-G.~Liu, J.~Zhang and Y.~-S.~Piao,
  Phys.\ Rev.\ D {\bf 84}, 063508 (2011)  [arXiv:1105.5713
  ]; 
  Z.-G.~Liu and Y.-S. Piao,
  Phys.\ Lett.\ B {\bf 718}, 734 (2013)  [arXiv:1207.2568 ].


\bibitem{Hinterbichler:2012fr}
  K.~Hinterbichler, A.~Joyce, J.~Khoury and G.~E.~J.~Miller,
  JCAP {\bf 1212}, 030 (2012)  [arXiv:1209.5742 [hep-th]]; 
  Phys.\ Rev.\ Lett.\  {\bf 110}, 24, 241303 (2013)  [arXiv:1212.3607
  [hep-th]].

\bibitem{Piao:2003ty}
  Y.~S.~Piao and E.~Zhou,
  Phys.\ Rev.\ D {\bf 68}, 083515 (2003)
  [hep-th/0308080].

\bibitem{Rubakov:2014jja}
  V.~A.~Rubakov,
  arXiv:1401.4024 [hep-th].

\bibitem{Khoury:2001wf}
  J.~Khoury, B.~A.~Ovrut, P.~J.~Steinhardt and N.~Turok,
  Phys.\ Rev.\ D {\bf 64}, 123522 (2001)  [hep-th/0103239]; 
  J.~-L.~Lehners and P.~J.~Steinhardt,
  Phys.\ Rev.\ D {\bf 87}, no. 12, 123533 (2013)  [arXiv:1304.3122 [astro-ph.CO]].

\bibitem{Piao:2003zm}
  Y.~S.~Piao, B.~Feng and X.~m.~Zhang,
  Phys.\ Rev.\ D {\bf 69}, 103520 (2004)
  [hep-th/0310206];
  Y.~S.~Piao,
  Phys.\ Rev.\ D {\bf 71}, 087301 (2005)
  [astro-ph/0502343].

\bibitem{Liu:2013kea}
  Z.~G.~Liu, Z.~K.~Guo and Y.~S.~Piao,
  Phys.\ Rev.\ D {\bf 88}, 063539 (2013)
  [arXiv:1304.6527].


\bibitem{Dudas:2012vv}
  E.~Dudas, N.~Kitazawa, S.~P.~Patil and A.~Sagnotti,
  JCAP {\bf 1205} (2012) 012; 
  N.~Kitazawa and A.~Sagnotti,
  arXiv:1402.1418 [hep-th].


\bibitem{Ramirez}
  E. Ramirez, D. J. Schwarz, Phys.\ Rev.\ D {\bf 85}, 103516
  (2012)  [arXiv:1111.7131 [astro-ph.CO]].

\bibitem{Borde:2001nh}
  A.~Borde, A.~H.~Guth and A.~Vilenkin,
  Phys.\ Rev.\ Lett.\  {\bf 90}, 151301 (2003)  [gr-qc/0110012].

\bibitem{Khoury:2010gw}
  J.~Khoury and G.~E.~J.~Miller,
  Phys.\ Rev.\ D {\bf 84}, 023511 (2011)  [arXiv:1012.0846
  [hep-th]]; 
  A.~Joyce and J.~Khoury,
  Phys.\ Rev.\ D {\bf 84}, 023508 (2011)  [arXiv:1104.4347 [hep-th]].

\bibitem{Geshnizjani:2011rm}
  G.~Geshnizjani, W.~H.~Kinney and A.~M.~Dizgah,
  JCAP {\bf 1202}, 015 (2012)  [arXiv:1110.4640 [astro-ph.CO]].


\bibitem{NRT}
  A.~Nicolis, R.~Rattazzi and E.~Trincherini,
  Phys.\ Rev.\ D {\bf 79}, 064036 (2009)  [arXiv:0811.2197 [hep-th]].


\bibitem{Piao:2010bi}
  Y.~-S.~Piao,
  Phys.\ Lett.\ B {\bf 701}, 526 (2011)  [arXiv:1012.2734
  ].


\bibitem{Rubakov:2009np}
  V.~A.~Rubakov,
  JCAP {\bf 0909}, 030 (2009)  [arXiv:0906.3693 [hep-th]].

\bibitem{Hinterbichler:2011qk}
  K.~Hinterbichler and J.~Khoury,
  JCAP {\bf 1204}, 023 (2012)  [arXiv:1106.1428 [hep-th]].

\bibitem{camb}
A.~Lewis, A.~Challinor and A.~Lasenby,
  Astrophys.\ J.\  {\bf 538}, 473 (2000)
  [astro-ph/9911177];
A.~Challinor and A.~Lewis,
  Phys.\ Rev.\ D {\bf 71}, 103010 (2005)
  [astro-ph/0502425].

\bibitem{Wang:2014abh}
  Y.~T.~Wang and Y.~S.~Piao,
  arXiv:1409.7153 [gr-qc].

\bibitem{Rubakov:2013kaa}
  V.~A.~Rubakov,
  Phys.\ Rev.\ D {\bf 88}, 044015 (2013)  [arXiv:1305.2614
  [hep-th]].

\bibitem{Liu:2013xt}
  Z.~-G.~Liu and Y.~-S.~Piao,
  Phys.\ Rev.\ D {\bf 88}, 043520 (2013)  [arXiv:1301.6833
  [gr-qc]].

\bibitem{Elder:2013gya}
  B.~Elder, A.~Joyce and J.~Khoury,
  Phys.\ Rev.\ D {\bf 89}, 044027 (2014)  [arXiv:1311.5889
  [hep-th]].

\bibitem{Liu:2013iha}
  Z.~G.~Liu, Z.~K.~Guo and Y.~S.~Piao,
  Eur.\ Phys.\ J.\ C {\bf 74}, 3006 (2014)  [arXiv:1311.1599 [astro-ph.CO]].


\bibitem{Piao:2004jg}
  Y.~-S.~Piao and Y.~-Z.~Zhang,
  Phys.\ Rev.\ D {\bf 70}, 043516 (2004)  [astro-ph/0403671].

\bibitem{Piao:2004me}
  Y.~-S.~Piao,
  Phys.\ Rev.\ D {\bf 70}, 101302 (2004)  [hep-th/0407258]; 
  Phys.\ Lett.\ B {\bf 677}, 1 (2009)  [arXiv:0901.2644 [gr-qc]].


\bibitem{Xia:2014tda}
  J.~Liu, Y.~-F.~Cai and H.~Li,
  J.\ Theor.\ Phys.\  {\bf 1}, 1 (2012)  [arXiv:1009.3372
  [astro-ph.CO]]; J.~-Q.~Xia, Y.~-F.~Cai, H.~Li and X.~Zhang,
  arXiv:1403.7623 [astro-ph.CO].

\bibitem{Qiu:2014nla}
  T.~Qiu,
  arXiv:1404.3060 [gr-qc].


\end{thebibliography}
\end{document}